\documentclass[aps,pre,twocolumn,longbibliography,showpacs,amsmath,amsmath,amssymb,amsfonts,notitlepage,superscriptaddress]{revtex4-1}

\usepackage{graphicx}% Include figure files
\usepackage{dcolumn}% Align table columns on decimal point
\usepackage{bm}% bold math
\usepackage{hyperref}% add hypertext capabilities
\usepackage{xcolor}
\usepackage{graphicx}
\usepackage{dcolumn}
\usepackage{bm}
\usepackage{color}
\usepackage{multirow}
\usepackage[normalem]{ulem}
\usepackage{color}
\usepackage{physics}
\usepackage{bbm}
\usepackage{tikz-cd}
\usepackage{tikz}
\usepackage{hyperref}

\begin{document}

\title{\textbf{Large deviations, effective dynamics, and dynamical phase transitions in random maps}
}%

\author{Ángela Tajuelo-Valbuena}
\email[]{atajuelo@correo.ugr.es}
\affiliation{Departamento de Electromagnetismo y F\'isica de la Materia, Universidad de Granada, Granada 18071, Spain}

\author{Jorge Hidalgo}
\email[]{jorge.hidalgo@uco.es}
\affiliation{Departamento de Física, Universidad de Córdoba, 14071 Córdoba, Spain}
\affiliation{Institute Carlos I for Theoretical and Computational Physics, Universidad de Granada, Granada 18071, Spain}

\author{ Ricardo Guti\'errez}
\email[]{ricardo.gutierrez@uc3m.es}
\affiliation{Universidad Carlos III de Madrid, Departamento de Matem\'aticas, Grupo Interdisciplinar de Sistemas Complejos (GISC), Avenida de la Universidad, 30 (edificio Sabatini), 28911 Leganés (Madrid), Spain}
\author{Carlos P\'erez-Espigares}
\email[]{carlosperez@ugr.es}
\affiliation{Departamento de Electromagnetismo y F\'isica de la Materia, Universidad de Granada, Granada 18071, Spain}
\affiliation{Institute Carlos I for Theoretical and Computational Physics, Universidad de Granada, Granada 18071, Spain}

%


\begin{abstract}
Large deviation theory provides a natural framework to characterize rare fluctuations of time-averaged observables in deterministic and stochastic systems. Building on previous results for deterministic chaotic maps, here we extend the Doob transform to random maps, constructing an effective dynamics in which rare events become typical. This makes it possible to directly reproduce the dynamics of rare trajectories that are exponentially difficult to sample under the original dynamics. We further apply spectral large-deviation methods to characterize dynamical phase transitions in random maps. The spectral structure of the tilted operator provides insight into the mechanisms underlying these transitions and, in particular, allows us to identify the appropriate initial states that select one dynamical phase or another. Our results extend the connection between large deviation theory, spectral methods, and effective dynamics to random chaotic maps, providing a general framework to study rare events and dynamical phase transitions in these systems.

\end{abstract}

\maketitle
\section{Introduction}
\label{sec:intro}

Rare or highly improbable events play a fundamental role in complex systems. Despite their low probability of occurrence, when they do take place, they can have a profound impact on the system. Examples include natural disasters, financial collapses, and extreme climate conditions \cite{Klioutchnikov2017,ragone2018,Galfi2021}. Moreover, rare fluctuations are crucial to understanding nonequilibrium systems. By characterizing the probabilities of such fluctuations, large deviation theory (LDT) provides a suitable statistical framework for extending concepts such as free energy and entropy to nonequilibrium settings, where dynamics, unlike in equilibrium thermodynamics, plays a decisive role \cite{oono89a,touchette09,garrahan09a,jack20b}.

This issue is particularly relevant in the study of chaotic systems. Since they are deterministic, fluctuations can only arise from variations in their initial conditions \cite{beck93,ott02,strogatzbook}. The question is then whether LDT can be applied to characterize the statistics of these fluctuations, as is routinely done for stochastic systems. A recent study successfully demonstrated that LDT can indeed be applied to chaotic maps \cite{smith22}, with the same degree of generality employed in the study of Markov chains or diffusion processes. This breakthrough not only made it possible to calculate the probabilities of highly improbable fluctuations but also opened the door to addressing another fundamental question: what is the underlying dynamics responsible for producing such fluctuations? The latter question was answered by extending the so called generalized Doob transform, well known in stochastic systems \cite{simon2009,popkov10a,jack2010,chetrite13a,chetrite15a}, to one-dimensional chaotic maps \cite{Gutierrez2023}. The resulting transformation provides an effective dynamics in which fluctuations that are exponentially unlikely under the original dynamics become typical. More recently, these results have been generalized to chaotic maps of higher dimensionality  \cite{yllari2025}.

A key question remains open, however, which is the central focus of this work: whether LDT can be applied, and the Doob transform extended, to a class of maps that combine stochastic and deterministic dynamics. These systems, known as random maps, evolve according to a set of possible deterministic maps, one of which is randomly selected at each time step  \cite{pelikan84,morita85,buzzi00,dajani03,gora03,bahsoun05,inoue2012,sato19,kalle22,yan24,monthus2025}. The maps typically represent simple stretching and contraction mechanisms, giving rise, nonetheless, to remarkably rich dynamical behavior, including nontrivial invariant measures \cite{kalle22,monthus2025}, intermittency and anomalous dynamics \cite{sato19,yan24}. Understanding fluctuations in these systems therefore requires accounting for both the chaotic evolution generated by the individual maps and the stochastic sequence of maps selected along each trajectory. 

\begin{figure}[t]
\includegraphics[width=\linewidth]{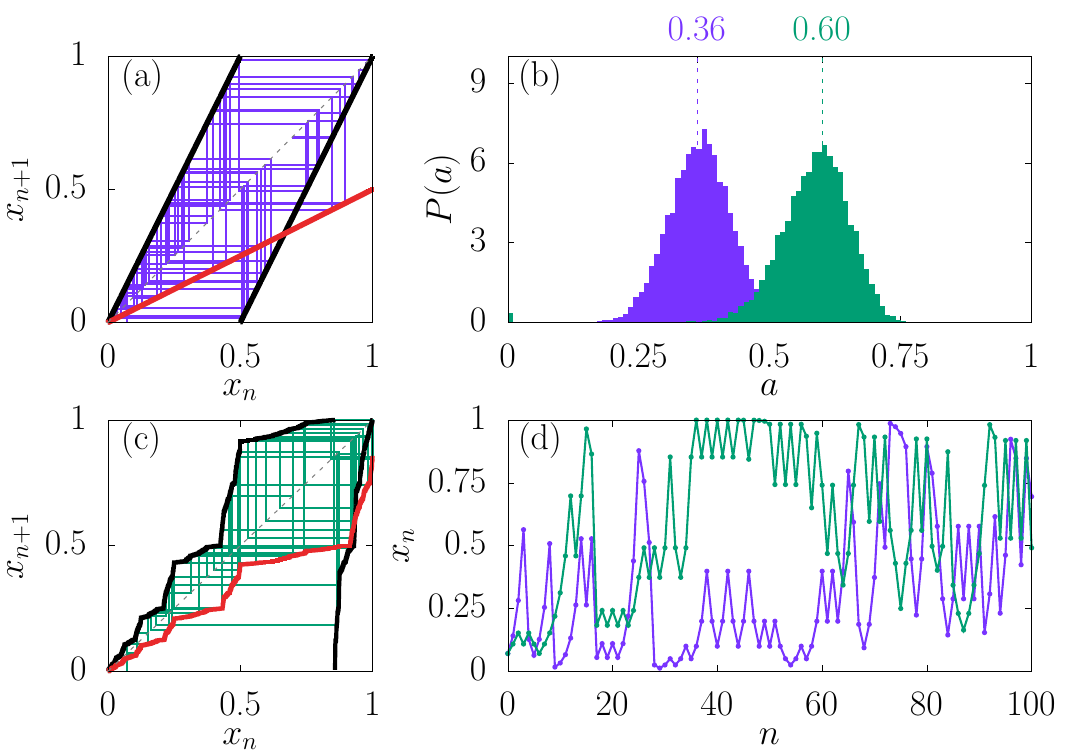}
\caption{(a) Cobweb plot of a trajectory for the Pelikan map (\ref{pelikan}) with $p=0.7$. (b) Histograms, $P(A_N=a)$, based on $10^{5}$ trajectories for the time-averaged observable $A_N=N^{-1}\sum_{n=0}^{N-1}x_n$. Purple corresponds to the Pelikan map and green to the Doob effective map for $s=-0.5$ displayed in panel (c). Their averages are, respectively, $\left<A\right>=0.36$ and $\left<A\right>_{s=-0.5}=0.60$. (c) Cobweb plot of a trajectory for the Doob effective map. (d) Representative trajectories obtained for the Pelikan map (purple) and the Doob effective map (green) for $N=200$ time steps. Results obtained with $N_x=3\times10^5$ cells (uniform mesh) and 30 power-method iterations.}
\label{fig:1D_v1.pdf}
\end{figure}

The first objective of this work is to extend the LDT framework and the Doob construction to random maps. While the theoretical development is general for time-averaged observables $A_N=N^{-1}\sum_{n=0}^{N-1}g(x_n)$ of any random map, we apply it to specific observables of the Pelikan map \cite{pelikan84}, a paradigmatic random dynamical system. This map is obtained by randomly alternating an expanding map, $f_1(x)=2 x\hspace{-1.5mm}\mod 1$ with probability $p$ and a contracting one, $f_2(x)=x/2$, with probability $1-p$, at each time step $n$, as illustrated in Fig.~\ref{fig:1D_v1.pdf}(a). We are interested in finding the Doob effective dynamics which reproduce a rare event of the the probability $P(A_N=a)$. See the purple histogram of Fig.~\ref{fig:1D_v1.pdf}(b) for an illustration based on the observable $A_N$ with local contribution $g(x)=x$, and the purple line in Fig.~\ref{fig:1D_v1.pdf}(d) for a representative trajectory. By means of the generalized Doob transform we unveil the effective random map, see Fig.~\ref{fig:1D_v1.pdf}(c), whose typical trajectories, as shown in green in Fig.~\ref{fig:1D_v1.pdf}(d), reproduce the rare events of the original system, as depicted in the green histogram of Fig.~\ref{fig:1D_v1.pdf}(b). The effective-dynamics approach previously developed for deterministic chaotic maps is thus generalized to systems in which deterministic and stochastic ingredients coexist.

The study of rare events in a variety of systems has revealed that the spectral approach to large deviations underlying the Doob transform, beyond providing an efficient representation of rare trajectories, offers a powerful framework to investigate dynamical phase transitions (DPTs) \cite{bodineau05a,garrahan07a,hurtado11a,perez-espigares13a,jack15a,tsobgni16a,lazarescu17a,perez-espigares18b,perez-espigares19a,gutierrez21b,banuls19a}. 
In stochastic many-body systems, the microscopic origin of DPTs can be understood from the spectral structure of tilted and Doob-transformed generators \cite{hurtadogutierrez23}, which has also been reported in deterministic chaotic maps \cite{Gutierrez2023,smith22}. 
Our large-deviation framework to random maps offers a new setting to explore the emergence of DPTs and investigate their underlying spectral mechanisms.
In the case of the Pelikan map \cite{pelikan84}, we uncover a dynamical phase diagram containing distinct trajectory phases separated by critical lines. These phases correspond to qualitatively different mechanisms by which atypical fluctuations are sustained and become accessible through the associated Doob dynamics.

A central result of our analysis is that these transitions admit a clear spectral interpretation. At the critical points, the dominant spectral gap of the tilted Frobenius-Perron operator closes, so that more than one eigenfunction contributes to the long-time Doob dynamics. As a consequence, the stationary state of the effective dynamics is no longer uniquely selected by the bias alone but can retain information about the initial probability distribution. The different dynamical phases can therefore be understood in terms of the structure of the dominant spectral subspace, providing a trajectory-space counterpart of phase coexistence. We complement this spectral picture with an analysis of the trajectories and time correlations close to the transition, thereby connecting the macroscopic (i.e.\! long-time-averaged) signatures of criticality with the microscopic dynamics responsible for the rare fluctuations.

The paper is organized as follows. In Sec. II we develop the general large-deviation framework for random maps. We introduce the tilted Frobenius-Perron operator for time-averaged observables and show how its leading spectral properties determine the statistics of rare fluctuations. This allows us to construct the generalized Doob transform and the corresponding effective random map that makes a prescribed rare event typical. In Sec. III we apply these ideas to the Pelikan map. After introducing the model and its relevant properties, we determine its dynamical phase diagram for an observable given by the state itself, we characterize the trajectories underlying the different phases, and we analyze the critical points through both time correlations and the spectrum of the tilted operator. Then we study the large deviations of another relevant observable, namely, the finite-time Lyapunov exponent (FTLE) of the Pelikan map. Finally, in Sec. IV we summarize our results and discuss possible extensions of the present framework.

\section{Rare events in random maps: theoretical framework}
\label{sec:theory}

\subsection{Random maps: definition and Frobenius-Perron operator}
\label{subsec:rm}

The random maps that we will consider are discrete-time systems whose evolution at time $n\in \mathbb{N}$ is given by
\begin{equation}
x_{n+1}=\begin{cases}
f_1(x_n),\ \text{with probability}\ p_1,\\
f_2(x_n),\ \text{with probability}\ p_2,\\
\cdots \\
f_m(x_n),\ \text{with probability}\ p_m.
\end{cases}
\label{map}
\end{equation}
Here, $m$ deterministic maps $f_j:I\to I$  ($j=1,2,\ldots, m$), defined on a compact interval of the real line $I$, are chosen with probability $0\leq p_j \leq 1$, satisfying $\sum_{j=1}^m p_j = 1$. In particular, the random map whose large deviations we will analyze numerically in detail in the next section is based on the simplest, most widely studied case of $m=2$ branches ---$f_1$ being a chaotic map, $f_2$ a regular map. See Fig.~\ref{fig:1D_v1.pdf} (a) for a cobweb plot, and Fig.~\ref{fig:1D_v1.pdf} (d) (purple line) for the corresponding trajectory.

We consider an ensemble of trajectories evolved with the random map \eqref{map} from the probability density of initial values $\alpha_0(x)$. The probability density at time $n$ follows
\begin{equation}
\alpha_{n+1}(x)=L[\alpha_n(x)],
\label{probevol}
\end{equation}
where $L$ is the Frobenius-Perron (FP) operator, defined as a generalization of the corresponding operator for deterministic maps (see, e.g., Ref.~\cite{beck93}), 
\begin{equation}
L[\alpha(x)]\!=\!\int_{I} dy\, \alpha(y)\! \sum_{j=1}^m p_j \delta(x-f_j(y))\!=\!\sum_{j=1}^m p_j\!\sum_{z_j}\frac{\alpha(z_j)}{|f_j'(z_j)|}
\label{FP}
\end{equation}
As indicated by the Dirac deltas, this is an integral over probabilities restricted to the preimages of $x$ under $f_j$ (with probability $p_j$), denoted collectively as $z_j$, which satisfy $f_j(z_j) = x$ ---if $f_j$ invertible, there will be just one preimage, $z_j = f_j^{-1}(x)$. The sum over $j=1,2,\cdots m$ accounts for the fact that the value $x$ may be reached by an application of each of the maps $f_j$ in Eq.~\eqref{map}, selected with its corresponding probability $p_j$. We are interested in cases for which the probability density converges at long times to the natural invariant density of the random map, denoted by $\rho(x)$, satisfying $\rho(x)=L[\rho(x)]$. 

Further, the adjoint FP operator (also known as Koopman operator) reads
\begin{equation}
L^\dagger[\alpha(x)]  = \sum_{j=1}^m p_j\, \alpha(f_j(x)).
\label{Koop}
\end{equation}
These definitions and others that ensue generalize those found for deterministic maps \cite{Gutierrez2023} by linearity.
For $\alpha(x)=\mathbbm{1}(x)$ (with $\mathbbm{1}(x)=1, \forall x\in I$), probability conservation, i.e., \! $\int L[\beta(x)] dx=\int \beta(x)dx=1$, implies that $L^\dagger[\mathbbm{1}(x)]=\mathbbm{1}(x)$ almost everywhere (a.e.), i.e.,~except perhaps in a set of Lebesgue measure zero. Hence, the invariant density $\rho(x)$ and $\mathbbm{1}(x)$ (together with functions equal to them a.e.) are eigenfunctions of the FP operator (\ref{FP}) and its adjoint (\ref{Koop}), respectively, both associated with the eigenvalue having the largest real part, which is $1$.

\subsection{Time-averaged observables and biased ensembles}
\label{subsec:sensemble}

Consider a trajectory of $N$ steps, $\omega_N = x_0 \to x_1\to \cdots \to x_{N-1}$ generated with the random map in Eq.~(\ref{map}). The probability of any such trajectory, which we denote as $P(\omega_N)$, is given by the density of initial conditions $\alpha_0(x)$ and the multinomial probability distribution with probabilities $\{p_j\}_{j=1}^m$ associated with a particular branching realization, according to which, e.g., $x_1 = f_2(x_0)$, $x_2 = f_3(x_1)$, $x_3 = f_7(x_2)$ and so on. This random element of the dynamics can be represented by a random variable $j_n$ at each step $n\in \mathbb{N}$. Thus $j_n = j \in \{1,2,\ldots, m\}$, if $f_j$ is chosen for the next step, $x_{n+1} = f_j(x_n)$, which occurs with probability $p_j$.

Denoting by $k_j \in \{0,1,2\ldots,N-1\}$ the number of transitions that are performed with map $f_j$, $k_j = \sum_{n=0}^{N-1} \delta_{j_n,j}$ (where $\delta_{j_n,j}$ is the Kronecker delta),
\begin{equation}
P(\omega_N) = \alpha_0(x_0)\, \prod_{j=1}^m p_j^{k_j},
\label{densitytraj}
\end{equation}
with $\sum_{j=1}^m k_j = N-1$. Notice that there are many trajectories with the same probability for the same initial value $x_0$, namely all $(N-1)!/k_1! k_2! \cdots k_m!$ branching realizations containing $k_j$ steps with map $f_j$ (irrespective of the order in which the different maps occur).

We focus on the statistics across trajectories of time-averaged observables,
\begin{equation}
A_N(\omega_N)=\frac{1}{N}\sum_{n=0}^{N-1}g(x_n),
\label{obs}
\end{equation}
where the scalar $g:I \to \mathbb{R}$, is the local contribution. Since the system is assumed to be ergodic, the long-time average of the observable $A_N$ converges to the ensemble average of $g(x)$ with respect to the invariant density $\rho(x)$ for generic trajectories:
\begin{equation}
\langle A \rangle = \lim_{N\to \infty} A_N(\omega_N) =\int_{I} dx\, \rho(x)  g(x).
\label{ergod}
\end{equation}

Yet the observable $A_N$ (\ref{obs}) is fluctuating, as its value depends on the trajectory $\omega_N$. The probability for the observable to take a value $a$ in a trajectory of $N$ steps is
\begin{equation}
P_N(A_N=a) = \int d\omega_N  P(\omega_N)\, \delta\left(A_N(\omega_N)-a\right)  \sim e^{-NI(a)}
\label{LDP}
\end{equation}
where integration with respect to $P(\omega_N)$ with measure $d\omega_N$ is shorthand for standard integration over the initial condition $x_0$ and summation from $k=0$ to $N-1$, see Eq.~(\ref{densitytraj}). The probability density, which from now on we simply denote as $P_N(a)$, is thus assumed to be dominated by the exponential form given in Eq.~(\ref{LDP}) for $N\gg 1$. The validity of this assumption, which is frequently made (as it leads to consistent results) in the analyses of rare events of stochastic systems of various kinds (see e.g.~Ref.~\cite{touchette09}), will be vindicated by our results below. This was also the case in the examples considered for discrete-time deterministic maps in Refs.~\cite{smith22,Gutierrez2023,yllari2025,Monthus2024,Defaveri2025,defaveri2026}.

The large deviation function (LDF) \cite{touchette09} $I(a)$ in Eq.~\eqref{LDP}, also known as the rate function, gives the rate at which the probability concentrates around $\langle A \rangle = \lim_{N\to \infty} \int da\, a\, P_N(a)$, see Eq.~(\ref{ergod}), as $N$ increases. This function is positive, $I(a)\ge 0$, and is equal zero at $\langle A \rangle$, fluctuations away from it becoming exponentially unlikely in time. Yet a direct estimation of the LDF $I(a)$ is challenging, so an alternative way to find relevant information about the distribution $P_N(a)$ \eqref{LDP} is frequently adopted. It is based on the moment-generating function
\begin{align}
Z_N(s) &= \int d\omega_N P(\omega_N) e^{-s N A_N(\omega_N)}\nonumber\\ 
&= \int   da P_N(a)\,e^{-s N a} \sim e^{N\theta(s)},
\label{zs1}
\end{align}
which also follows a large-deviation principle, in this case given by the scaled cumulant generating function (SCGF) $\theta(s)$, whose derivatives are proportional to the cumulants of $A_N$ \cite{touchette09}. From \eqref{LDP} and \eqref{zs1}, applying a saddle-point approximation for long times $N\gg 1$, the SCGF arises as the Legendre transform of the LDF,
\begin{equation}
\theta(s)=-\min_{a}[I(a)+sa].
\label{leg}
\end{equation}
Indeed, $I(a)$ and $\theta(s)$  play roles that are analogous to those of the entropy and the Helmholtz free energy, respectively, in equilibrium statistical mechanics.  As for $Z_N(s)$ (\ref{zs1}), it is a dynamical partition function that normalizes the exponentially-tilted distribution
\begin{equation}
P_{N,s}(a)=\frac{e^{-s N a} P_N(a)}{Z_N(s)},
\label{tiltdist}
\end{equation}
which biases the original distribution towards atypical fluctuations through the conjugate parameter $s$. The latter acts like an inverse temperature, but it can take either positive values (favoring values of $A$ smaller than the unbiased average $\langle A\rangle_{s=0} = \langle A \rangle$) or negative values (favoring values greater than $\langle A\rangle$).

The tilting of the distribution that results in $P_{N,s}(a)$ \eqref{tiltdist} thus amounts to a change from an (microcanonical) ensemble of trajectories   with fixed $A_N = a$, to a new (canonical) ensemble of trajectories ---the $s$-ensemble \eqref{tiltdist}---, where $A_N$ fluctuates across trajectories \cite{garrahan09a}, with an average value
\begin{equation}
\langle A \rangle_s = \lim_{N\to \infty} \int da\, a\, P_{N,s}(a)
\label{ave_s}
\end{equation}
that is fixed by the choice of $s$. Thus ``flat'' averages over initial conditions $\alpha_0$ and branching realizations $\{k_j\}$, cf.~Eq.~(\ref{densitytraj}) and Eq.~\eqref{LDP}, are replaced by exponentially-weighted averages of the form $e^{-s N a}$, cf.~Eq.~(\ref{tiltdist}). For a given fluctuation $a$, the appropriate choice of $s$ is the one that yields it as the average value, i.e.,\! $\langle A \rangle_s = a$. In terms of the SCGF and the LDF, this corresponds to $-\theta'(s)=a$ and $I'(a)=-s$, respectively \cite{touchette09}.
\subsection{Tilted Frobenius-Perron operator, SCGF}
\label{subsec:tilted op}

Following the argument in  Ref.~\cite{Gutierrez2023} for deterministic maps, one can show that the dynamical partition function $Z_N(s)$, Eq.~\eqref{zs1}, can be expressed as
\begin{equation}
Z_N(s)=\int_{I}dx\, L_s^N[\alpha_0(x)],
\label{zs2}
\end{equation}
where the $N$-th order composition, $L_s^N[\alpha_0(x)] = L_s[L_s[L_s[\cdots L_s[L_s[\alpha_0(x)]]\cdots]]]$, is performed with the tilted FP operator $L_s$ \cite{smith22},
\begin{align}
L_s[\alpha(x)] &=\!\int_{I} dy\, \alpha(y) \!\left[\sum_{j=1}^m p_j   e^{-s g(y)} \delta(x\!-\!f_j(y))\right]\nonumber\\
&= \sum_{j=1}^m \left(p_j\!\sum_{z_j}\frac{\alpha(z_j) e^{-s g(z_j)}}{|f_j'(z_j)|}\right).
\label{tiltop}
\end{align}
Here, the sum over $j$ accounts for the different branches, and the sum over $z_j$ accounts for the (in principle) various preimages of $x$ under the map $f_j$. Moreover, the adjoint of the tilted FP operator is given by
\begin{equation}
L^{\dagger}_s[\alpha(x)]=\sum_{j=1}^m p_j\,   e^{-s g(x)}  \alpha[f_j(x)].
\label{tiltadjop}
\end{equation}
Note that Eq.~(\ref{tiltop}) reduces to Eq.~(\ref{FP}), and Eq.~(\ref{tiltadjop}) reduces to Eq.~(\ref{Koop}), in the absence of tilting, i.e.~for $s=0$.

It is straightforward to check, by focusing on the spectrum of the tilted FP operator $L_s$, that for long times $Z_N(s)$ follows the large deviation form $Z_N(s)\sim e^{N\theta(s)}$ [in agreement with Eq.~\eqref{zs1}] where the exponential of the SCGF \eqref{leg}, $e^{\theta(s)}$, is the eigenvalue having the largest real part of $L_s$ and $L_s^{\dagger}$,
\begin{align}
L_s[r_s(x)] &= e^{\theta(s)} r_s(x),\nonumber \\
L^{\dagger}_s[l_s(x)] &= e^{\theta(s)} l_s(x).
\label{eigprob}
\end{align}
Here $r_s(x)$ and $l_s(x)$ are the right and left eigenfunctions, respectively, normalized in such a way that $\int_{I}dx\, r_s(x)\,l_s(x)=1$ and $\int_{I} dx\, r_s(x)=1$. (The degenerate case where these eigenfunctions are not unique will be considered below.)
Finding the LDF $I(a)$ is possible by solving the eigenvalue problem \eqref{eigprob}, with $s$ as a parameter, and using SCGF $\theta(s)$ in the inverse of the Legendre transformation \eqref{leg}. Moreover, once $\theta(s)$ is known, all cumulants of $A_N$ are known for the corresponding $s$. Hence, one can adjust the mean of the distribution to a rare fluctuation, $\langle A \rangle_s=a$, by choosing $s$ so that $-\theta'(s)=a$ [equivalently, $-s=I'(a)$].

The focus of this work lies in the trajectories that sustain such a fluctuation, as our aim is to find another random map of the form (\ref{map}) which generates them naturally in its stationary dynamics, yielding a long-time average of the observable equal to the $s$-ensemble average  $\langle A \rangle_s$. The latter can be written in terms of the eigenfunctions as
\begin{equation}
\langle A_N \rangle_s \xrightarrow[N\to\infty]{\hspace{0.8cm}}\int_{I} dx\, g(x)\, l_s(x)\, r_s(x).
\label{midstat3}
\end{equation}
See Ref.~\cite{yllari2025} for a derivation in the context of chaotic maps.  We conclude that a random map that generates trajectories corresponding to the $s$-ensemble (\ref{tiltdist}) must have as natural invariant density
\begin{equation}
\rho_s(x)=l_s(x)\,r_s(x),
\label{sinv}
\end{equation}
which reduces to $\rho(x)$ for the unbiased dynamics $s=0$, yielding (\ref{ergod}).

\subsection{Generalized Doob transform and Doob map}
\label{subsec:doob}

As an intermediate step in our way to finding the effective map that naturally generates the rare trajectories of interest as typical events, we derive a FP operator having the invariant density (\ref{sinv}). Following Ref.~\cite{Gutierrez2023}, we introduce the so-called Doob operator, which for random maps generalizes in a straightforward manner to
\begin{equation}
L_{s}^D[\alpha(x)] = e^{-\theta(s)} l_{s}(x) L_{s}[\left(l_{s}(x)\right)^{-1}\alpha(x)].
\label{LD}
\end{equation}
This is a FP operator that satisfies the desired conditions: (i) it is a proper probability-preserving generator, $(L_{s}^D)^{\dagger}[\mathbbm{1}(x)]=\mathbbm{1}(x)$, which the tilted operator \eqref{tiltop} is not, and (ii) its invariant density is $\rho_{s}(x)$, $L_{s}^D[\rho_{s}(x)] = \rho_{s}(x)$. The latter implies that it generates the ensemble of trajectories yielding the tilted distribution \eqref{tiltdist} in the long-time limit, as shown in detail in the Supplemental Material of Ref.~\cite{Gutierrez2023}.

The Doob generator \eqref{LD}, as a FP operator, acts on probability densities $\alpha(x)$, yet we are ultimately interested in finding a Doob effective map. This is a random map that, starting from an arbitrary initial condition $x_0$, typically generates the trajectories that sustain the rare events corresponding to $s \neq 0$. Such Doob effective map is based on a transformation  $\gamma_{s}$ of the state $x$ taken over (typical) trajectories of the original map ---which follow a distribution $\rho(x)$ for long times, see Eq.~\eqref{ergod}---, 
so that the long-time distribution of $\tilde{x}=\gamma_{s}(x)$ is given by  $\rho_{s}(\tilde{x})$ (\ref{sinv}).

In the present large-deviation framework for random maps, as happens in the deterministic case \cite{Gutierrez2023,yllari2025}, the Doob effective map is topologically conjugate to the original dynamics, in contrast to the stochastic case where the Doob dynamics corresponds to a reweighing of the original (unbiased) transition probabilities between configurations \cite{jack2010,chetrite15b}.  Because topological conjugation preserves ergodicity, the long-time average of the observable in the transformed (rare) trajectories converges to its $s$-ensemble average  [i.e.,~the average taken with respect to the invariant density $\rho_{s}(\tilde{x})$],
\begin{equation}
\langle A \rangle_{s}\!=\!\lim_{N\to \infty}\frac{1}{N}\sum_{n=0}^{N-1}g(\tilde{x}_n)\!=\!\int_{I}d\tilde{x}\, g(\tilde{x})\rho_{s}(\tilde{x}),
\label{ave_s2}
\end{equation}
cf.~Eq.~(\ref{ergod}).

Following \cite{Gutierrez2023}, finding $\tilde{x}=\gamma_{s}(x)$ is analogous to transforming values \( x \) of a random variable \( X \), which is distributed according to \( \rho(x) \), into values \( \tilde{x} \) of a random variable \( \tilde{X} \), which follows the distribution \( \rho_{s}(\tilde{x}) \), by applying the inverse transformation method \cite{devroye06a}. In the original dynamics $x \in I\subset \mathbb{R}$, where $I$ is some compact interval (i.e.,~the phase space of the map $f$), and it turns out that we also have $\tilde{x} \in I$ by construction, since its density $\rho_{s}(\tilde{x})$ is obtained from the eigenfunctions in Eq.~(\ref{eigprob}). In fact, $\gamma_{s}$ is a bijection on $I$, which we take to be increasing as in \cite{Gutierrez2023} ---the procedure can be alternatively carried out with a decreasing transformation.

Assuming that $\rho(x)$ and $\rho_{s}(\tilde{x})$ are integrable and strictly positive \footnote{While the original invariant density $\rho(x)$ satisfies these properties in the cases that we consider, $\rho_{s}(\tilde{x})$ (which is obtained from the eigenfunctions of the tilted Frobenius-Perron operator) displays singular behavior for some deterministic maps \cite{Monthus2024}. This has a limited effect on our results, given their approximate nature based on phase-space discretizations, yet it may be relevant when considering exact results.}, the cumulative distributions $F(x) = \int^x du\, \rho(u)$ and $F^D_{s}(\tilde{x}) = \int^{\tilde{x}} du\, \rho_{s}(u)$ are continuous and increasing (hence invertible) \footnote{Whenever we omit the lower (upper) limit in one-dimensional integrals, they are integrated from the minimum (maximum) of their range, e.g.,~if a function is defined in $[a,b]$, then $\int^x$ will be an integration over $[a,x]$.}, then we can write
\begin{align}
F^D_{s}(\tilde{x})&=P(\tilde{X}\leq \tilde{x})=P(\gamma_{s}(X)\leq \tilde{x})\nonumber\\
&=P(X\le \gamma_{s}^{-1}(\tilde{x}))=F[\gamma_{s}^{-1}(\tilde{x})].
\end{align}
Therefore $F^D_{s}(\tilde{x})=F(x)$ implies that
\begin{equation}
\tilde{x}=\gamma_{s}(x) = (F^D_{s})^{-1}[F(x)],
\label{g1d}
\end{equation}
yielding the desired transformation, $\gamma_s = (F_s^D)^{-1}\circ F$.

\section{Rare events in random maps: Dynamical Phase Transitions of the Pelikan Map}
\label{sec:pelikan}

\subsection{Pelikan map: definition and relevant properties}
\label{subsec:pelikan}

In this section we study the large deviations of the Pelikan map, a paradigmatic example of a random map that was introduced four decades ago \cite{pelikan84}. It comprises two randomly-selected deterministic branches, an expanding map with rate $2$ ---perhaps the simplest chaotic system, known as the doubling map---, and a contracting map with rate $1/2$, both defined over the interval $I = [0,1]$:
\begin{equation}
x_{n+1} = \begin{cases}
f_1(x_n) = 2 x_n \bmod 1, & \text{with probability } p,\\ 
f_2(x_n) = x_n/2, & \text{with probability } q.
\end{cases}
\label{pelikan}
\end{equation}
This conforms to the scheme given in Eq.~\eqref{map}, with $m=2$ branches, and probabilities $p_1 = p$ and $p_2 = q$; as $q=1-p$, we will frequently focus on $p$ as the randomness parameter. A typical trajectory arising from this map is provided in Fig.~\ref{fig:1D_v1.pdf} as an illustration, both in a cobweb representation, see Fig.~\ref{fig:1D_v1.pdf}(a), and in a standard ($x_n$ vs $n$) representation, see purple line in Fig.~\ref{fig:1D_v1.pdf}(d).

The FP operator \eqref{FP} for the Pelikan map is
\begin{equation}
L[\alpha(x)] = \frac{p}{2} \left[\alpha\!\left(\frac{x}{2}\right)+\alpha\!\left(\frac{x+1}{2}\right)\right] +  2 q\, \alpha(2 x)\, \chi_{[0,1/2]} (x).
\label{FP_pelikan}
\end{equation}
Here, $x/2$ and $(x+1)/2$ are the preimages of $x$ under the doubling map $f_1$, and $2x$ is the only preimage of $x$ under the contracting map $f_2$, which only exists for $x \in [0,1/2]$. The latter is ensured by the presence of the indicator function $\chi_{[0,1/2]} (x)$, defined to be one on $[0,1/2]$, vanishing elsewhere.

While the invariant measure of the expanding branch $f_1(x) = 2 x \bmod 1$, which has a positive FTLE, is uniform, that of the contracting branch $f_2(x) = x/2$, which has a negative FTLE, is a delta at the origin. Their combination in the Pelikan map \eqref{pelikan} yields an invariant density with a more complex structure when the dynamics is chaotic and the invariant measure is absolutely continuous, which happens for $p>p^*$, where $p^*=1/2$ is a critical sampling probability \cite{pelikan84}. Moreover, in Ref.~\cite{yan24}, the Pelikan map has been shown to exhibit a non-trivial transition from chaotic to regular dynamics at $p^*$, defined by a zero FTLE, and characterized by long-time anomalous dynamics,  where correlations decay following a power law ---while they do so exponentially for other values of $p$.

\subsection{Dynamical phase transitions}
\label{subsec:DPT}

\subsubsection{Large-deviation formalism for $g(x) = x$}
\label{subsubsec:LD-x}

We first study the observable $A_N = N^{-1} \sum_{n=0}^{N-1} x_n$, which corresponds to the general form \eqref{obs}, with $g(x) = x$. The tilted FP operator \eqref{tiltop} for the Pelikan map and this observable results from an exponential tilting of the terms in the FP operator \eqref{FP_pelikan}:
\begin{align}
L_s[\alpha(x)] &= \frac{p}{2} \left[e^{-s\frac{x}{2}} \alpha\!\left(\frac{x}{2}\right) +e^{-s \frac{x+1}{2}}\alpha\!\left(\frac{x+1}{2}\right)\right]\nonumber\\ 
&+  2 q\, e^{-s 2 x} \alpha(2 x)\, \chi_{[0,1/2]} (x).
\label{LsPel}
\end{align}
As for the adjoint tilted FP operator \eqref{tiltadjop}, it is
\begin{equation}
L_s^\dagger[\alpha(x)] = e^{-s x}\left(p\, \alpha(2 x \bmod 1) + q\, \alpha(x/2) \right).
\label{LsAdjPel}
\end{equation}

The leading eigenvalue and eigenfunctions ara obtained numerically with the power method \cite{Gutierrez2023}: $L_s$ and $L_s^\dagger$ are applied iteratively to an initial (e.g., constant) function, normalizing at each step, until convergence to $r_s(x)$ and $l_s(x)$, respectively, is achieved, while the normalization factors converge to the largest eigenvalue $e^{\theta(s)}$. A detailed description of the procedure can be found in the Supplemental Material of Ref.~\cite{Gutierrez2023}. When subleading eigenfunctions are also needed, we use instead the Arnoldi method~\cite{lehoucq98}, which computes the few leading eigenvalues without obtaining the whole spectrum, except when dealing with matrices of moderate size, for which we  use full matrix diagonalization. Analytical results for the right eigenfunction $r_s(x)$ allow us to confirm the accuracy of our numerical estimates; see Appendix~\ref{app:righteig}.

\begin{figure}[t]
\includegraphics[width=\linewidth]{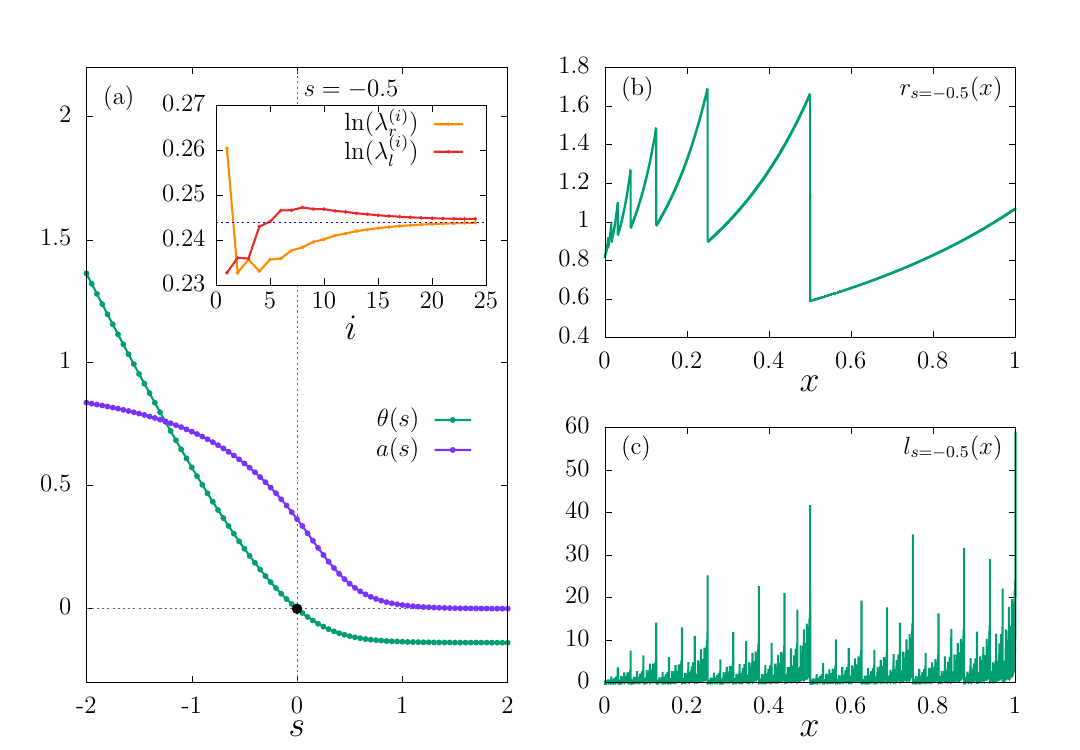}
\caption{Results for the time-averaged observable $A_N=N^{-1}\sum_{n=0}^{N-1}x_n$ for the Pelikan map with $p=0.7$ and $s\in [-2,2]$. (a) The green line shows the result obtained with the numerical method for the SCGF $\theta(s)$. The purple line corresponds to $a=-\theta'(s)$. The inset shows the convergence of the logarithm of the largest eigenvalue $(\lambda(s) = e^{\theta(s)})$ associated with the right and left eigenfunctions using the numerical method. (b) Numerical result for the right eigenfunction with $s=-0.5$. (c) Numerical result for the left eigenfunction with $s=-0.5$.}
\label{fig:2C_v1.pdf}
\end{figure}

In Fig.~\ref{fig:2C_v1.pdf}(a) we show the SCGF $\theta(s)$ and the mean value of the observable $a(s) \equiv \langle A\rangle_s = -\theta'(s)$ for $p=0.7$, which corresponds to chaotic dynamics in the natural ($s=0$) evolution. The inset shows the convergence of the logarithm of the largest eigenvalue $\lambda(s)= e^{\theta(s)}$ associated with the left and right eigenfunctions as a function of the number of iterations of the power method. The eigenfunctions themselves are shown for $s=-0.5$ in Fig.~\ref{fig:2C_v1.pdf}(b), in the case of the right-eigenfunction $r_{s=-0.5}(x)$, and Fig.~\ref{fig:2C_v1.pdf}(c), in the case of the left-eigenfunction $l_{s=-0.5}(x)$. These results were obtained by discretizing the interval $I=[0,1]$ into $N_x = 3\times 10^5$ uniformly spaced cells, with spacing $\Delta x = N_x^{-1}$, and performing 30 iterations of the power method.

\subsubsection{Phase diagram}
\label{subsubsec:corrs}

\begin{figure}[t]
\includegraphics[width=\linewidth]{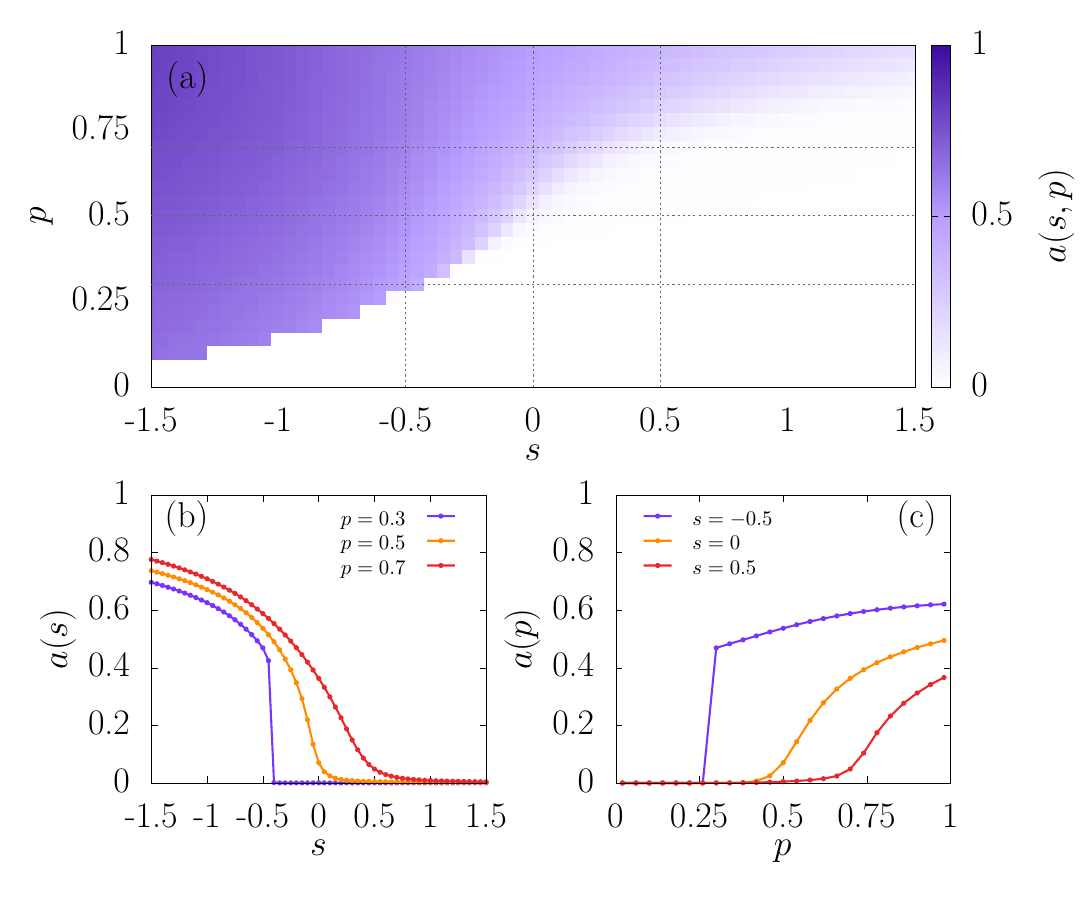}
\caption{ (a) Phase diagram of $a(s,p)$ for observable $g(x)=x$. (b) Dependence of $a(s)$ on $s$ for selected values of $p$. (c) Dependence of $a(s)$ on $p$ for selected values of $s$. Panels (b) and (c) illustrate the continuous or discontinuous nature of the transition. Results obtained with $N_x=1.5\times 10^4$ (uniform mesh), Arnoldi method.}
\label{fig:phasediagram-x}
\end{figure}

A similar analysis is performed for range of values of $p$ spanning from $p=0$ (deterministic dynamics given by the contracting branch $f_2$) to $p=1$ (deterministic dynamics given by the expanding branch $f_1$), passing through many points where the random nature of the map is relevant. The average value of the observable $\langle A\rangle_s$ as a function of $s$ but also of $p$, denoted $a(s,p)$, is displayed in Fig.~\ref{fig:phasediagram-x}(a). This is a dynamical phase diagram with control parameters given by the sampling probability $p$ and the tilting parameter $s$, and reveals the existence of DPTs in the Pelikan map for observable $g(x) = x$. Indeed, for a given $p$, there is a transition from positive $a(s,p)$ to an inactive region where $a(s,p) \approx 0$ as  $s$ increases, the specific $s$ that gives the transition point being an increasing function of $p$. Moreover, the nature of the transition appears to be different for $p<0.5$ and $p>0.5$. 

In fact, Fig.~\ref{fig:phasediagram-x}(b), showing $a(s,p)$ as a function of $s$ for fixed $p$, clearly illustrates how the continuous phase transition at $p^*=0.5$ in the natural dynamics ($s=0$), studied in detail in Ref.~\cite{yan24}, is displaced towards $s>0$ for $p>0.5$, while becomes displaced towards $s<0$ for $p<0.5$, in the latter case seemingly displaying the discontinuous character of a first-order DPT. A complementary view is provided in Fig.~\ref{fig:phasediagram-x}(c), showing $a(s,p)$ as a function of $p$ for fixed $s$. In the following, we will denote as $p^*$ the critical point where the transition takes place for a given $s$.

\subsubsection{Critical points: characterization via time correlations}
\label{subsubsec:corrs}
The nature of the DPTs found in the Pelikan map for the observable given by $g(x)=x$ can be also characterized through the temporal correlations of the rare trajectories. For a state observable $h(x)$, we consider the connected autocorrelation in the stationary regime, $C_s(n)=\langle h(x_0)h(x_n)\rangle_s-\langle h\rangle_s^2$. Here, $h(x)$ is the local contribution of an observable $B_N = N^{-1} \sum_{n=0}^{N-1} h(x_n)$, which in principle could be different from the observable used in the tilting. As shown in Appendix~\ref{app:corr} using the adjoint Doob operator, one can write
\begin{equation}
\langle h(x_0)\,h(x_n)\rangle_s
= \int_I dx\,h(x)\,\left(L_s^D\right)^n\big[h(x)\,\rho_s(x)\big].
\label{eq:corrDoob}
\end{equation}
That is, the correlation is obtained by evolving $n$ steps with the Doob generator  the observable weighted by the density, $h(x) \rho_s(x)$, and then integrating the result with $h(x)$.

To quantify how fast correlations decay, we define a characteristic correlation length as the first moment of $|C_s(n)|$,
\begin{equation}
\xi_s=\frac{\sum_{n=0}^{n_{\max}} n\,|C_s(n)|}{\sum_{n=0}^{n_{\max}}|C_s(n)|},
\label{eq:xi}
\end{equation}
where the absolute value accounts for possible oscillations of $C_s(n)$, and the cutoff $n_{\max}$ becomes relevant if the first moment diverges. For exponential decay, the correlation length $\xi_s$ defined in \eqref{eq:xi} (with $n_{\max}\to \infty$) corresponds to the characteristic decay time: $C_s(n) \sim e^{-n/\xi_s}$ for $\xi_s\gg 1$.

\begin{figure}[t]
\includegraphics[width=\linewidth]{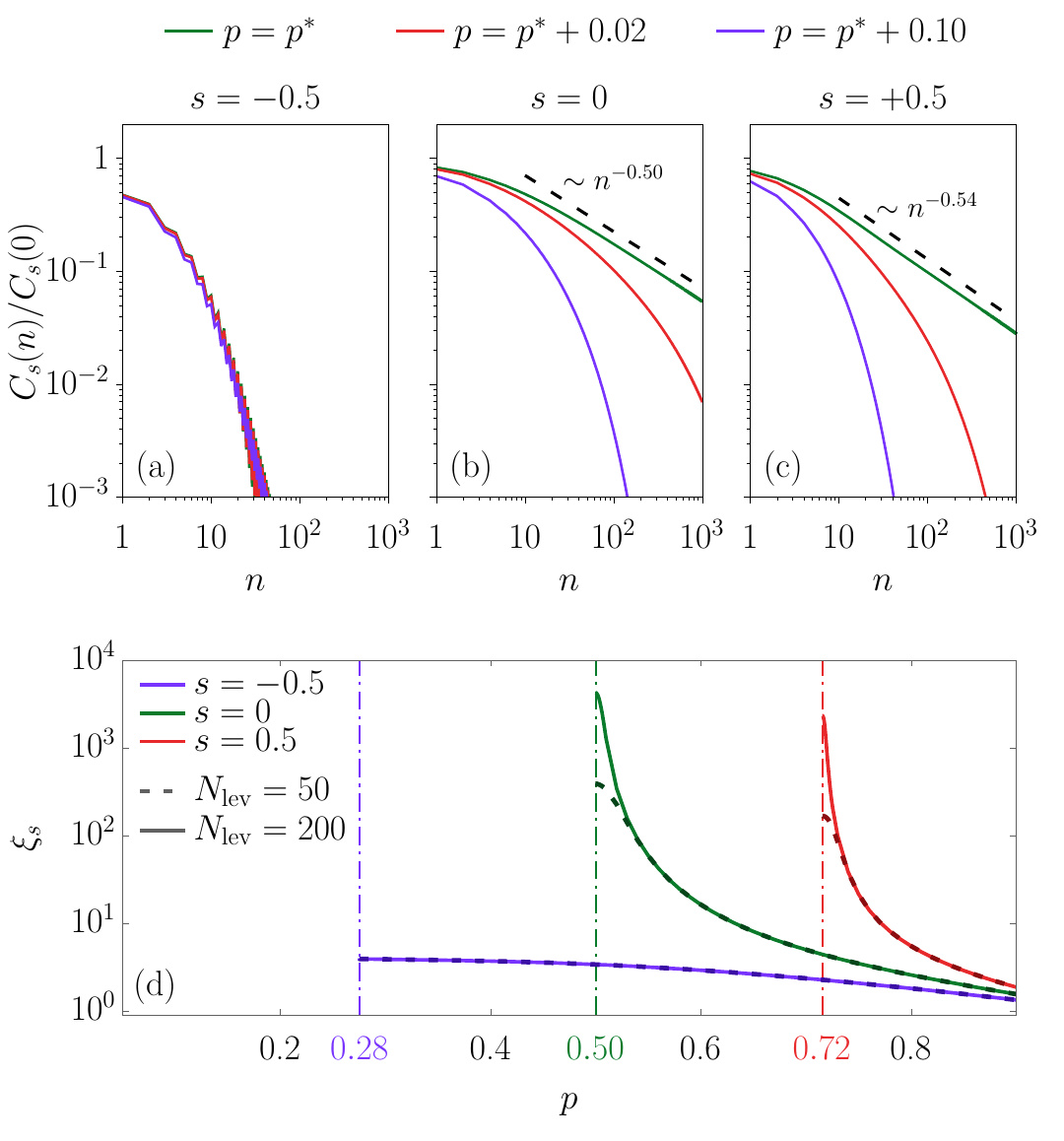}
\caption{Autocorrelation of the biased dynamics (observable $g(x)=x$) for $s=-0.5$, $0$ and $0.5$. (Upper panels) Normalized correlation $C_s(n)/C_s(0)$ in log-log scale at $p=p^*(s)$, $p^*(s)+0.02$ and $p^*(s)+0.10$. Dashed lines are power-law fits at $p=p^*(s)$ over $[10, 10^3]$. Parameters: $N_\text{lev}=800$ and $p^*=0.28$, $0.5$ and $0.72$, respectively. (Lower panel) Correlation length $\xi$ (first moment of $|C_s(n)|$) as a function of $p$ for two values of $N_\text{lev}$. Vertical lines mark $p^*(s)$, located as the maximum of $\xi$. Sums are truncated at $n_\text{max}=2\times10^4$.
}
\label{fig:correlations}
\end{figure}

To compute Eq.~\eqref{eq:corrDoob} for a given tilting $s$ and probability $p$, we need the eigenfunctions $l_s(x)$ and $r_s(x)$ and the SCGF $\theta(s)$ (see Appendix~\ref{app:corr} for details), which we again obtain with the power method. Close to the critical points, however, we find that the uniform mesh, which we employed for the discretization of the interval $I = [0,1]$ in  previously-displayed results, fails to accurately compute time correlations. To overcome this problem we use a logarithmic mesh, better suited to the structure of the invariant density of the Pelikan map (i.e.,~the right eigenfunction for $s=0$) found in Ref.~\cite{pelikan84} and, more generally, to the right eigenfunction for $s\neq 0$ (see Appendix~\ref{app:righteig}). Its resolution near the origin is set by the number $N_{\rm lev}$ of levels, i.e., intervals $[2^{-(\ell+1)},2^{-\ell})$, it resolves, and details of its implementation are given in Appendix~\ref{app:logmesh}.

Figure~\ref{fig:correlations} shows the correlation function $C_s(n)$ for the observable $h(x)=g(x)=x$ at $s=-0.5$, $s=0$, and $s=0.5$
[Figs.~\ref{fig:correlations} (a)–(c), respectively], with $p\geq p^*(s)$, and the corresponding characteristic length $\xi_s$ [see Fig.~\ref{fig:correlations}(d)]. For $s=-0.5$, correlations always decay within a few steps, with a zigzag between even and odd $n$; $\xi_s$ remains bounded at $p^*$ and independent of $N_{\rm lev}$. This is consistent with a discontinuous transition. For $s=0$ and $s=0.5$, instead, $C_s(n)\propto n^{-\beta}$ at $p^*$, with exponents compatible with the exact value $\beta=1/2$ obtained by Monthus \cite{monthus2025} for the unbiased ($s=0$) Pelikan map at its critical point $p^*=1/2$ \footnote{In Ref.~\cite{monthus2025}, the correlation is computed with the initial condition $x_0$ drawn from a uniform distribution, rather than from the stationary density. At $p^*=1/2$, however, a similar behavior is expected for large time lags.}, which corresponds to an unbiased random walk between levels. Correspondingly, $\xi_s(p)$ peaks sharply at $p^*$, with a height that grows with $N_{\rm lev}$, suggesting a divergence in the continuum limit, as expected for a continuous transition. 

Finally, it is worth noting that, for $s=0.5$, the fitted exponent $\beta\simeq0.54$ appears to deviate slightly from $1/2$. This deviation may reflect the finite fitting window or the uncertainty in locating $p^*$, and a more detailed analysis would be needed to determine whether the exponent for $s>0$ genuinely differs from $1/2$.

\subsubsection{Critical points: characterization via spectral analysis}
\label{subsubsec:spect}

Spectral signatures provide valuable insight into DPTs \cite{hurtadogutierrez23}. In what follows we shall describe how the spectral signatures of the Pelikan map behave at the previously-identified critical points, where a gap closure in the spectrum of the tilted FP operator \eqref{LsPel} will be shown to take place. As we need eigenvalues and eigenfunctions other than the leading one, we extend the notation of Subsec.~\ref{subsec:tilted op} introduced in the context of the eigenproblem \eqref{eigprob}. Specifically we denote the eigenvalues and eigenfunctions of order $j$ as $e^{\theta_j(s)}$, $r_s^j(x)$ and $l_s^j(x)$, where $\theta_0(s) \geq \theta_1(s)\geq \theta_2(s)\geq \cdots$. In this notation, $e^{\theta_0(s)}$, $r_s^0(x)$ and $l_s^0(x)$ correspond to what we previously denoted simply as $e^{\theta(s)}$, $r_s(x)$ and $l_s(x)$; while $e^{\theta_1(s)}$, $r_s^1(x)$ and $l_s^1(x)$ in turn correspond to the first subleading eigenvalue and corresponding eigenfunctions. We will omit the dependence of $\theta_j(s)$ on $s$ for notational compactness. 

In our study of the spectral signatures of critical points, we start by focusing on the evolution of probability densities in the effective (Doob) dynamics:  
\begin{equation}
    \alpha_n(x)=(L_s^D)^n[\alpha_0(x)]=e^{-n \theta_0
}l_s^0(x)L_s^n[(l_s^0(x))^{-1}\alpha_0(x)].
\label{eq:LDn}
\end{equation}
We will be particularly interested in the stationary regime of such dynamics. To gain insight into the dominant contributions for $n\gg 1$, we must take into account the spectral decomposition of the tilted operator,
\begin{equation}
   L_s[\alpha(x)]=\sum_{j=0}^m e^{\theta_j} r_s^j(x)\!\int dy\, l_s^j(y) \alpha(y)
   \label{eq:Ls-spec}
\end{equation}
Here, $m$ is in practice given by the discretization whereby $L_s$ is approximated by a matrix, but we do not particularize its value as the following discussion centers on the lowest terms in the decomposition. Combining Eqs.~\eqref{eq:LDn} and \eqref{eq:Ls-spec},
\begin{align}
    \alpha_n(x)&=l_s^0(x)r_s^0(x)\label{eq:spec-evol}\\
    &+l_s^0(x)\sum_{j\ge 1}^m e^{-n(\theta_0-\theta_j)} r_s^j(x)\int dy\, l_s^j(y)(l_s^0(y))^{-1}\alpha_0(y)\nonumber
\end{align}
Each term with $j\geq 1$ decays at a rate set by the spectral gap $\theta_0-\theta_j \geq 0$. In particular, for nondegenerate spectra, i.e.~$\theta_0 - \theta_1 > 0$, all the subleading terms vanish and the stationary state is unique, namely $l_s^0(x) r_s^0(x)$. Hence, a DPT requires an emergent degeneracy of the leading eigenspace, i.e.~a closure of the spectral gap.
If the tilted FP operator presents spectral degeneracy or level touching we get that $\theta_0=\theta_1$, and then in the stationary state, i.e. for long times $n\gg1$, we obtain,
\begin{equation}
    \alpha_\text{stat}(x)=l_s^0(x)r_s^0(x)+l_s^0(x) r_s^1(x)\int l_s^1(y)(l_s^0(y))^{-1}\alpha_0(y)dy.
    \label{eq:stat-deg}
\end{equation}

Depending on the initial condition, one might naively expect the stationary state to be given by $l_s^0(x)r_s^0(x)$, $l_s^0(x)r_s^1(x)$, or a linear combination thereof. However, the biorthogonality relation $\int dx\, l_s^i(x)r_s^j(x)=\delta_{ij}$ ensures that $l_s^0(x)r_s^0(x)$ is normalized, whereas $l_s^0(x)r_s^1(x)$ has zero integral and therefore cannot represent a probability density on its own. Moreover, the latter contribution must change sign unless it vanishes identically. Physical stationary states must therefore be constructed as normalized, non-negative linear combinations of these two contributions. We choose the normalization of the subleading eigenfunctions such that the two extremal stationary probability densities take the form
\begin{equation}
    \alpha_{\pm}(x)=l_s^0(x)r_s^0(x)\pm l_s^0(x)r_s^1(x),
\end{equation}
with $\alpha_\pm(x)\geq 0$ for all $x\in I$.

To clarify the role of the initial condition, let us first consider an initial density belonging to the stationary subspace,
\begin{equation}
    \alpha_0(x)=b\alpha_+(x)+(1-b)\alpha_-(x),
    \qquad 0\leq b\leq 1.
    \label{eq:alpha_0_deg}
\end{equation}
Using the biorthogonality relation, we immediately obtain
\begin{equation}
    \alpha_\text{stat}(x)=l_s^0(x)r_s^0(x)
    +(2b-1)l_s^0(x)r_s^1(x)
    =\alpha_0(x).
    \label{eq:stat-deg-2}
\end{equation}
Thus, at the critical point, the stationary state is no longer unique but depends on the initial probability density. In particular, $b=1$ and $b=0$ select the extremal stationary states $\alpha_+(x)$ and $\alpha_-(x)$, respectively, intermediate values of $b$ yielding convex combinations thereof.

Crucially, this dependence on the initial condition persists even when $\alpha_0(x)$ does not belong to the stationary subspace and contains contributions from higher-order eigenfunctions. Indeed, provided that $\theta_0=\theta_1>\theta_2$, all contributions with $j\geq 2$ decay exponentially with time, as shown in Eq.~\eqref{eq:spec-evol}, leaving a stationary state determined by the projection of the initial density onto the degenerate leading eigenspace.

\begin{figure}[t]
\includegraphics[width=\linewidth]{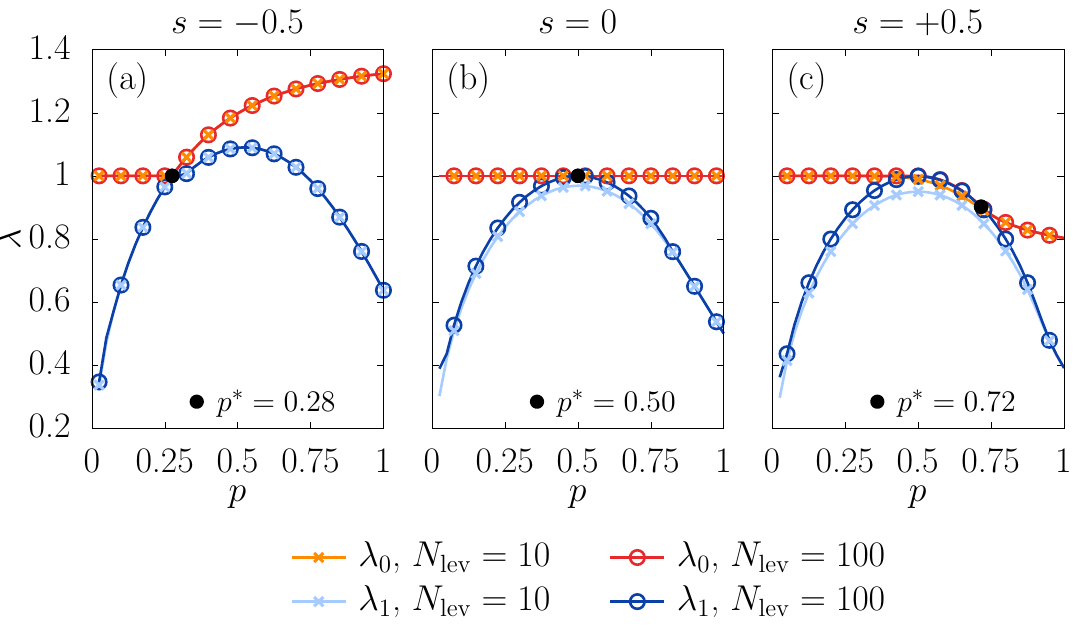}
\caption{The two leading eigenvalues, $\lambda_0$ and $\lambda_1$, of the operator $L_s$ as a function of $p$ for different values of $s$, computed with a logarithmic discretization based on $N_\text{lev}=10$ or $N_\text{lev}=100$ (see legend underneath). Results corresponding to (a) $s=-0.5$, (b) $s=0$ and (c) $s=+0.5$. In each panel, the black dot marks $p^*$ operationally defined as the value of $p$ at which the gap $|\lambda_0-\lambda_1|$ is smallest. Results obtained by full matrix diagonalization.}
\label{fig:eigenvalues}
\end{figure}

In Fig.~\ref{fig:eigenvalues} we show the two leading eigenvalues of $L_s$, denoted $\lambda_0(s) = e^{\theta_0(s)}$ and $\lambda_1(s) = e^{\theta_1(s)}$, as a function of $p$. On the logarithmic mesh, the moderate size of the matrix representing $L_s$ allows us to obtain both eigenvalues by full numerical diagonalization. Their behavior depends strongly on the sign of $s$. For $s<0$, two eigenvalue branches cross at $p^*$, so that the leading eigenvalue switches abruptly from one branch to the other, which is compatible with a discontinuous transition. For $s>0$, instead, the second eigenvalue approaches the leading one smoothly, and the two remain nearly degenerate over a whole interval of $p$ below $p^*$, where the gap decreases as the resolution of the mesh is increased. Both behaviors are contemplated in Ref.~\cite{hurtadogutierrez23} for stochastic systems, and their regime of applicability switches at the natural dynamics, $s=0$, in conformity with the conclusions derived from the phase diagram in Fig.~\ref{fig:phasediagram-x}, and the nature of the correlation decay in Fig.~\ref{fig:correlations}.

\begin{figure}[t]
\includegraphics[width=\linewidth]{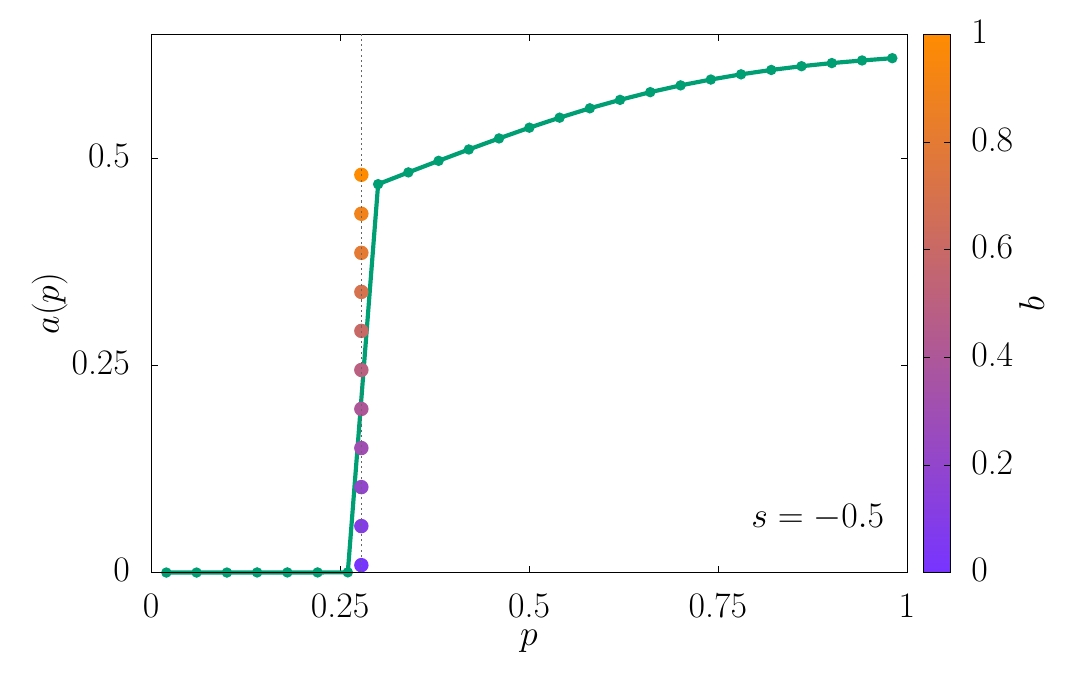}
\caption{Biased observable $a(p)$ for $s=-0.5$ (green points connected with solid line). Points colored with a gradient from purple to yellow represent the ensemble average $\int_0^1 x\alpha_{\mathrm{stat}}(x)dx$, evaluated at the critical point $p^*\approx 0.28$ for the family of stationary states $\alpha_{\mathrm{stat}}(x)=b\alpha_+(x)+(1-b)\alpha_-(x)$, with $0\leq b\leq 1$. Results obtained with $N_x=1.5\times10^4$ cells (uniform mesh), Arnoldi method.}
\label{fig:spectral-gradient}
\end{figure}

Fig.~\ref{fig:spectral-gradient} illustrates the biased observable $a(p)$ for $g(x)=x$ as a function of $p$ for $s=-0.5$, calculated as the minus derivative of the SCGF, i.e., $a(p)=-\theta'(s)$ (green points). Points colored along a gradient from purple to yellow represent the ensemble average $\int_0^1 x\alpha_{\mathrm{stat}}(x)dx$, evaluated at the critical point $p^*\approx 0.28$ [where a level crossing takes place, as shown in Fig.~\ref{fig:eigenvalues}(a)] for the family of stationary states $\alpha_{\mathrm{stat}}(x)=b\alpha_+(x)+(1-b)\alpha_-(x)$, with $0\leq b\leq 1$. The abrupt change at $p= p^*$ indicates that the vanishing gap allows two distinct stationary states to coexist. The extreme values of $b$ in Eq.~\eqref{eq:alpha_0_deg} continuously connect with the values observed on either side of the transition: $b=0$ (purple point) gives $a(p) \approx 0$, as observed for $p<p^*$, and $b=1$ (yellow point) gives $a(p) \approx 0.5$, close to value of the branch that emanates for $p>p^*$. Intermediate values of $b$ in $(0,1)$ monotonically connect these possibilities, showing coexistence with different weights. These results highlight the dependence of the stationary state on the initial distribution at the gap closure point, and offer compelling evidence of the first-order nature of the transition for $s<0$.

\subsection{Finite-time Lyapunov exponent (FTLE)}
\label{subsec:lyap}

Our theoretical framework can in principle be extended to observables that depend not just on the state of the system $x_n$, with local contribution $g(x_n)$, but also on the branch of the random map that is chosen, with local contribution $g(x_n,j_n)$. Among them, perhaps the most important one is the FTLE, which is given by $g(x_n,j_n) = \ln|f_{j_n}'(x_n)|$. In the case of the Pelikan map, as the two deterministic maps involved have constant derivatives, the dependence is effectively on the branch $j_n\in \{1,2\}$ alone: $g(x_n,1) = \ln|f_1'(x_n)|=\ln 2$ and $g(x_n,2) = \ln|f_2'(x_n)| = -\ln 2$. 

The tilted FP operator \eqref{tiltop} for the Pelikan map and the FTLE is
\begin{align}
L_s[\alpha(x)] &= 2^{-(s+1)} p\left[\alpha\!\left(\frac{x}{2}\right) + \alpha\!\left(\frac{x+1}{2}\right)\right]\nonumber\\ 
&+  2^{s+1} q\, \alpha(2 x)\, \chi_{[0,1/2]} (x).
\label{LsPelLyap}
\end{align}
As previously done for $g(x)=x$, the right eigenfunction $r_s(x)$ is numerically obtained by applying the power-method to the right eigenproblem \eqref{eigprob} with this operator $L_s$. Quite unlike what happened for the previous observable, however, in the case of the FTLE the exact SCGF $\theta(s)$ and left eigenfunction $l_s(x)$ are easily obtained, as will be shown next. In fact, the right eigenfunction $r_s(x)$ can also be derived analytically, since the tilted FP operator in Eq.~\eqref{LsPelLyap} takes the same form ---up to a normalization factor shown to be $e^{\theta(s)}$ below--- as the original FP in Eq.~\eqref{FP_pelikan}, only with $p$ and $q$ replaced by $p_s = 2^{-s} p/(2^{-s} p + 2^{s} q)$ and $q_s = 1-p_s = 2^s q/(2^{-s} p + 2^{s} q)$, respectively, and the stationary density of \eqref{FP_pelikan} is known \cite{pelikan84, yan24}. Nonetheless, the advantage of using the analytical solution is unclear, as it is defined by a recurrence that must truncated numerically for the graphical representation of results.

Since the adjoint tilted FP operator \eqref{tiltadjop} is
\begin{equation}
L_s^\dagger[\alpha(x)] = 2^{-s} p\, \alpha(2 x \bmod 1) + 2^{s} q\, \alpha(x/2),
\label{LsAdjPelLyap}
\end{equation}
the left eigenproblem \eqref{eigprob} becomes
\begin{equation}
2^{-s} p\, l_s(2 x \bmod 1) + 2^{s} q\, l_s(x/2) = e^{\theta(s)}l_s(x).
\label{lefteiglyap}
\end{equation}
To solve it, let us denote $M = \sup_{x\in I} l_s(x)$. Then the left-hand side of \eqref{lefteiglyap} is bounded by
\begin{equation}
2^{-s} p\, l_s(2 x \bmod 1) + 2^{s} q\, l_s(x/2) \leq (2^{-s} p + 2^s q) M,
\end{equation}
but it turns out that the upper bound is attained for the constant left eigenfunction $l_s(x) = M$, yielding as (logarithm of) the largest eigenvalue $\theta(s) = \ln(2^{-s} p + 2^s q)$. 
The precise value of $M$ is given by the normalization, $\int_I dx\,  r_s(x)\, l_s(x) = M \int_I dx\, r_s(x) = 1$ and $\int_I dx\, r_s(x) = 1$, yielding $M=1$. Hence, $l_s(x) = \mathbbm{1}(x)$ for all $s$ in this case. As a consequence, the Doob operator \eqref{LD}, based on the tilted FP operator \eqref{LsPelLyap}, is 
\begin{align}
L_{s}^D[\alpha(x)] &= e^{-\theta(s)} L_{s}[\alpha(x)] = \frac{1}{2^{-s} p + 2^s q} L_s [\alpha(x)]\\ &=  p_s\!\left[\alpha\!\left(\frac{x}{2}\right)\!+\!\alpha\!\left(\frac{x+1}{2}\right)\right] +  q_s \alpha(2 x)\, \chi_{[0,1/2]} (x).\nonumber
\end{align}
This further confirms that the effective dynamics is the natural ($s=0$) dynamics, given by the FP operator \eqref{FP_pelikan}, for rescaled probabilities $p_s$ and $q_s$.

The large-deviation statistics of the FTLE were previously derived by Monthus \cite{monthus2025} from the binomial distribution of expanding and contracting iterations. The SCGF obtained here is consistent with those results. Our focus, however, is on constructing the corresponding Doob effective dynamics and showing that it retains the structure of the original random map, with suitably modified probabilities.

\begin{figure}[t]
\includegraphics[width=\linewidth]{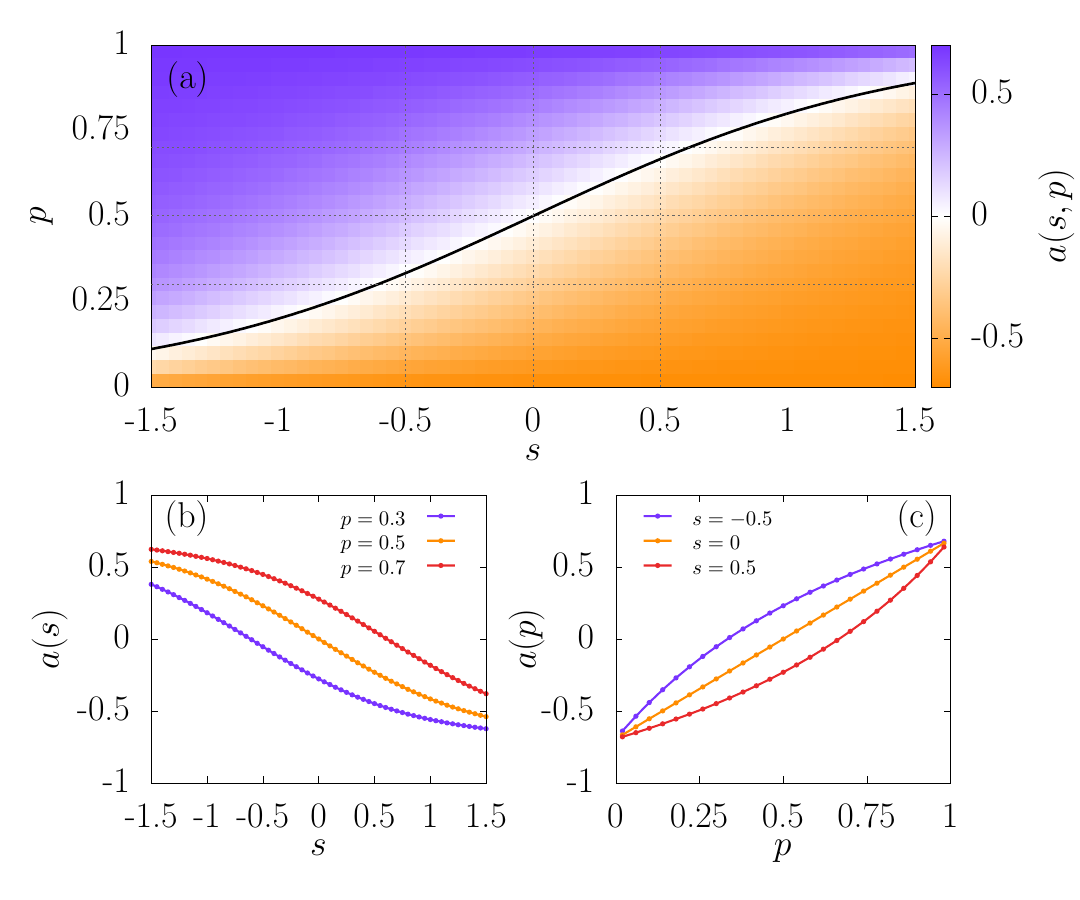}
\caption{(a) Phase diagram of $a(s,p)$, using the FTLE as observable. (b) Dependence of $a(s)$ on $s$ for selected values of $p$. (c) Dependence of $a(s)$ on $p$ for selected values of $s$. Results obtained with $N_x= 1.5\times 10^4$ cells (uniform mesh), Arnoldi method.}
\label{fig:phasediagram-lyapunov}
\end{figure}

The phase diagram of the FTLE, giving its average value as a function of the two parameters $a(s,p)$, is shown in Fig.~\ref{fig:phasediagram-lyapunov}(a). In this case, $a(s,p)$ has been computed exactly as (minus) the first derivative of the SCGF $\theta(s)$
\begin{equation}
a(s,p) = -\frac{\partial}{\partial s} \ln(2^{-s} p + 2^s q) =  \frac{2^{-s} p - 2^s q}{2^{-s} p + 2^s q}\ln 2.
\label{asp_lyap}
\end{equation}
In terms of the rescaled probabilities, this is $a(s,p) = (p_s - q_s) \ln 2$. The black continuous line indicates the point where the FTLE vanishes, i.e. where $p_s = q_s$, yielding
\begin{equation}
p^* = \dfrac{2^s}{2^s + 2^{-s}},
\end{equation}
separating a dynamics dominated by (chaotic) expansion from one dominated by contraction. Notice that $p^* = 1/2$ for $s=0$, as expected, since in the natural dynamics with equiprobable branches the $\ln 2$ contribution of the expanding branch and the $-\ln 2$ contribution of the contracting branch precisely cancel. Moreover, $p^*$ it is an increasing function of $s$ bounded between two horizontal asymptotes, namely, $p\to 1$  as $s\to \infty$, and $p\to 0$  as  $s\to -\infty$. This can be understood as follows: when the bias strongly favors small FTLE ($s\gg 0$) the expanding branch needs to be much more likely than the contracting one ($p$ must be very close to $1$) to avoid having a negative FTLE, while for biases strongly in favor of large FTLE ($s\ll 0$) a very small probability for the expanding branch ($p$ slightly above $0$) suffices to obtain a positive FTLE. 

Fig.~\ref{fig:phasediagram-lyapunov}(b) and (c) show the average $a(s,p)$ for fixed $p$ and a function of $s$, and for fixed $s$ as a function of $p$, respectively. In this case, all changes are smooth across the phase diagram, as expected from the analytical dependence in Eq.~\eqref{asp_lyap}.

\section{Conclusions}

We have presented a versatile framework for making rare events typical in random maps. Those rare events correspond to unlikely fluctuations of time-averaged observables. They are explored and selected through a biasing (``tilting'') of the dynamics, and the analysis of large-deviation functions. A rare event of choice is made typical through a generalization of the procedure known as the generalized Doob transform of stochastic systems, which has been recently adapted to deterministic maps. Indeed, one of the main contributions of this paper is to extend this methodology to the case of random maps. 

Our framework is put in practice in the study of the Pelikan map, possibly the most widely-studied random map. We numerically unveil a line of critical points in the phase diagram of the position operator $g(x) = x$, which include both continuous and first-order DPTs. Their nature is investigated via an analysis of time-correlations and spectral signatures, which provide compelling and consistent evidence allowing us to characterize in great detail the nature of the critical points. Finally, we study the Pelikan map with a different observable, namely the FTLE. Besides its intrinsic importance in the qualitative understanding of a map that combines chaotic and contracting dynamics, this allows us to test our framework in the case of observables that depend on the choice of branch in the random map. Moreover, the fluctuations of the FTLE can be partially analyzed through analytical means, which complement and provide support to our numerical findings.

With this contribution, the large deviation framework that was developed for stochastic systems and recently adapted to deterministic (chaotic maps) \cite{smith22, Gutierrez2023, Monthus2024, yllari2025}, has been extended to the realm of random maps, which naturally combine stochasticity and determinism.

\begin{acknowledgments}

The research leading to these results has received funding from the I+D+i Grants PID2023-149365NB-I00, PID2023-149174NB-I00, and PID2024-159024NB-C21 by the Spanish Ministry and Agencia Estatal de Investigación MICIU/AEI/10.13039/501100011033 and EDRF funds (European Union). We are also grateful for the computing resources and technical support provided by PROTEUS, the supercomputing center of the Institute Carlos I for Theoretical and Computational Physics in Granada, Spain.

\end{acknowledgments}

\appendix

\section{Recurrence relation for the right eigenfunction $r_s(x)$ of the Pelikan map with observable $g(x) =x$}
\label{app:righteig}

The tilted FP operator of the Pelikan map for the observable with local contribution $g(x)=x$ is given in Eq.~\eqref{LsPel}. The right eigenproblem $L_s[r_s(x)] = e^{\theta(s)} r_s(x)$, in terms of $\tilde{r}_s(x) = e^{-s x} r_s(x)$ for convenience, becomes 
\begin{equation}
\tilde{r}_s\!\left(\frac{x}{2}\right)\!=\!\frac{2}{p} e^{\theta(s)\!+\!s x} \tilde{r}_s(x) -\frac{ 4 q}{p}\, \tilde{r}_s(2 x) \chi_{[0,1/2]} (x)-  \tilde{r}_s\!\left(\frac{x+1}{2}\right), 
\label{pelikanEq}
\end{equation}
after rearranging terms. Following the reasoning used to obtain the invariant density of the Pelikan map in Ref.~\cite{pelikan84}, corresponding to $s=0$, as well as compelling numerical evidence for other values of $s$ [see, e.g., Fig.~\ref{fig:2C_v1.pdf}(b)], we assume that $r_s(x)$ is discontinuous at $x = 2^{-\ell}$ for any positive integer $\ell$. We thus take $\tilde{r}_s(x)$ to be a piecewise defined function on the dyadic intervals bounded by such discontinuities, which also correspond to the levels in the logarithmic mesh employed for the study of correlations (see Appendix~\ref{app:logmesh}):
\begin{equation}
\tilde{r}_s(x) = \begin{cases}
\tilde{r}_s^{(0)}(x)\ \ x \in I_0 = \displaystyle \left[1/2,1\right),\\
\tilde{r}_s^{(1)}(x)\ \ x \in I_1 = \displaystyle \left[1/4,1/2\right),\\
\ \cdots\\
\tilde{r}_s^{(\ell)}(x)\ \ x \in I_\ell = \displaystyle \left[2^{-(\ell+1)},2^{-\ell}\right),\\
\ \cdots
\end{cases}
\label{eq:r_s_Rec}
\end{equation}
For the first step in the recurrence, Eq.~(\ref{pelikanEq}) yields
\begin{equation}
\tilde{r}_s^{(1)}\left(\frac{x}{2}\right) = \frac{2}{p} e^{\theta(s) + s x}\, \tilde{r}_s^{(0)}(x) - \tilde{r}_s^{(0)}\left(\frac{x+1}{2}\right),
\label{rec1}
\end{equation}
while for $\ell\geq 1$,
\begin{align}
\tilde{r}_s^{(\ell+1)}\!\left(\frac{x}{2^{\ell+1}}\right) &= \frac{2}{p} e^{\theta(s) + s \frac{x}{2^\ell}}\, \tilde{r}_s^{(\ell)}\!\left(\frac{x}{2^\ell}\right)\label{rec2}\\  
&-  \frac{ 4 q}{p}\, \tilde{r}_s^{(\ell-1)}\!\left(\frac{x}{2^{\ell-1}}\right) -   \tilde{r}_s^{(0)}\!\left(\frac{x}{2^{\ell+1}}+\frac{1}{2}\right),\nonumber
\end{align}
both expressions given in terms of $x \in I_0$.

\begin{figure}[h]
\includegraphics[width=\linewidth]{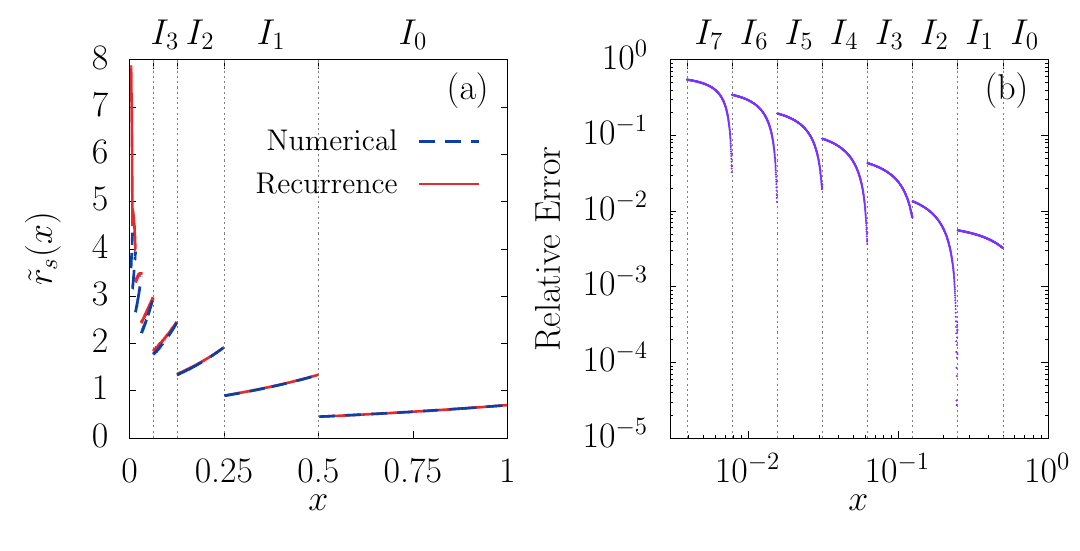}
\caption{(a) Numerical ($\tilde{r}^{\mathrm{num}}_s$) and recurrence ($\tilde{r}^{\mathrm{re}}_s$) right eigenfunction results for $s=-0.2$ and $p=0.6$, using a uniform mesh ($N_x=3\times10^5$) with 30 iterations of the power method. (b) Relative error, $\frac{|\tilde{r}^{\mathrm{num}}_s-\tilde{r}^{\mathrm{re}}_s|}{|\tilde{r}^{\mathrm{re}}_s|}$, between both methods, on a log-log scale. In both panels, the intervals $I_\ell$ are marked. }
\label{fig:recurrence}
\end{figure}

One can obtain $\tilde{r}_s^{(0)}(x)$ numerically, and then use the recurrence relation in Eqs.~\eqref{rec1} and \eqref{rec2} to check the consistency and accuracy of numerical results. In Fig.~\ref{fig:recurrence}(a) we show a comparison, for $s=-0.2$ and $p=0.6$, between the numerically-obtained (rescaled) right eigenfunction $\tilde{r}_s(x)$, which in the following we denote as $\tilde{r}^{\mathrm{num}}_s(x)$, and the one that results from taking the numerically-obtained $\tilde{r}_s^{(0)}(x)$ as input and obtaining $\tilde{r}_s^{(\ell)}(x)$ for $\ell\geq 1$ using this recurrence, denoted as $\tilde{r}^{\mathrm{re}}_s(x)$. The agreement is excellent in the first few iterations while some small discrepancies arise for smaller intervals close to the origin (i.e.\! for $I_\ell$ with sufficiently large $\ell$). Fig.~\ref{fig:recurrence}(b) shows the relative error between the two methods of computing $\tilde{r}_s(x)$, namely $|\tilde{r}^{\mathrm{num}}_s(x)-\tilde{r}^{\mathrm{re}}_s(x)|/|\tilde{r}^{\mathrm{re}}_s(x)|$, which shows a good agreement throughout the several intervals $I_\ell$ inspected, with the error increasing with larger $\ell$.

The results discussed in the main text for the Pelikan map with observable $g(x) = x$ are based on the numerically-obtained $r_s(x)$, which in the notation used in this Appendix corresponds to $r_s(x) = e^{s x}\,\tilde{r}^{\mathrm{num}}_s(x)$.

\section{Time correlations in the Doob dynamics}
\label{app:corr}

To address the computation of time correlations, it is important to clarify first that time-evolving averages can be equivalently computed by evolving probability densities or by evolving observables.  Thus, to compute the average of observable $h(x)$ at step $n$ in the original ($s=0$) dynamics, we may integrate $h(x)$ with respect to $\alpha_n(x) = L^n[\alpha_0(x)]$, which is the probability density evolved with the FP operator $L$ from the initial density $\alpha_0(x)$. But, equivalently, one can let the observable $h(x)$ evolve under the  adjoint (Koopman) operator  $L^\dagger$ an integrate with respect to the initial density:
\begin{equation}
\langle h(x_n) \rangle =  \int_I dx\, L^n[\alpha_0(x)]  h(x) = \int_I dx\, \alpha_0(x) (L^{\dagger})^n[ h(x)], 
\end{equation}
since $(L^n)^\dagger = (L^{\dagger})^n$. This equivalence between evolving densities with the FP operator and evolving observables with the adjoint operator is analogous to that between evolving states in Schr\"odinger's picture and evolving operators in Heisenberg's picture of quantum mechanics.

In the effective dynamics under consideration, for $s\neq 0$, the evolution is given by the Doob FP operator $L_s^D$, for densities, or its adjoint  $(L_s^D)^{\dagger}$, for observables, and the initial density is the stationary density $\rho_s(x)$. The two-point observable of interest is $h(x_0) h(x_n)$, where $h(x_n)$ is the $n$-fold evolution of $h(x_0)$. The latter, as it corresponds to an observable not a density, must be given by the adjoint operator $(L_s^D)^\dagger$. By reversing the argument above, we find
\begin{align}
\langle h(x_0) h(x_n) \rangle &=  \int_I dx\, \rho_s(x)\, h(x)\, ((L_s^D)^\dagger)^n[h(x)]\nonumber\\ 
&= \int_I dx\, (L_s^D)^n[\rho_s(x)\, h(x)]\, h(x),
\end{align}
which is just Eq.~\eqref{eq:corrDoob} introduced in Subsec.~\ref{subsubsec:corrs}. Using Eq.~\eqref{eq:LDn}, with $\theta_0=\theta(s)$ and $l_s^0=l_s$, we can reexpress this equation terms of the tilted operator $L_s$
%which follows from iterating the definition of the Doob generator,
%Eq.~\eqref{LD}: the factor $l_s(x)$ produced by each iteration cancels the
%factor $(l_s(x))^{-1}$ introduced by the next one,
applied on the weighted density
$\rho_s(x)\, h(x) = l_s(x)\, r_s(x)\, h(x)$. The $l_s(x)$ factors inside $L_s^n$ cancel out,
yielding
\begin{equation}
\langle h(x_0)\,h(x_n)\rangle_s = e^{-n\theta(s)} \int_I dx\, l_s(x)\, h(x)\,
L_s^n\big[h(x)\, r_s(x)\big].
\label{eq:corrtilted}
\end{equation}

This way of expressing Eq.~\eqref{eq:corrDoob} has some numerical advantages: (i) it does not require building a new operator, since
$L_s$ is already used in the power method to obtain $\theta(s)$, $l_s(x)$ and
$r_s(x)$; and (ii) it avoids dividing by $l_s(x)$, which might lead to amplified numerical error when $l_s(x)$ takes very small values.

\section{Logarithmic mesh}
\label{app:logmesh}

Close to $x=0$, the dynamics of the Pelikan map is best described by the level $\ell$ of the interval $I_\ell=[2^{-(\ell+1)},2^{-\ell})$ containing $x$, i.e., $\ell\simeq-\log_2 x$. The contracting map moves the state from level $\ell$ to
$\ell+1$, and the expanding map from $\ell$ to $\ell-1$ for $\ell\geq1$, while from $I_0$ it reinjects states into the whole interval $I$. A uniform mesh of $N_x$ equal cells only resolves about $\log_2 N_x$ levels, which artificially limits the walk between levels and makes correlations decay too fast.

The logarithmic mesh follows these levels down to $\ell=N_{\rm lev}-1$, so that the number of levels $N_{\rm lev}$ sets the smallest resolved scale, $2^{-N_{\rm lev}}$. In the upper levels the intervals $I_\ell$ are wide, so the tilting factor $e^{-sg(x)}$ and the observable can change significantly within a
single interval. These levels, that we take as $\ell<N_{\rm fine}$, are therefore divided into $M_{\rm sub}$ cells of equal width. In the deeper levels, $N_{\rm fine}\leq \ell< N_{\rm lev}$, the width $2^{-(\ell+1)}$ is so small that
these quantities that depend on $x$ are essentially constant within each interval, and a single cell per level is enough. A last cell $[0,2^{-N_{\rm lev}})$ represents $x\simeq0$. In our numerical implementations of the logarithmic mesh, we use
$N_{\rm fine}=12$ and $M_{\rm sub}=20$.

%\bibliography{refsLDRandomMaps}
\bibliography{referenciasDoobChaoticOct26}

\end{document}